\documentclass[12pt]{article}
\usepackage{a4,latexsym,pslatex,epsf,pictex,graphicx,times,psfrag,amssymb}
\usepackage[numbers]{natbibn}

\newcommand{\dd}[1]{{\rm{d}}{#1}}

\title{Three-dimensional bubble clusters: shape, packing and growth-rate}
\author{{\bf{S.J. Cox{$^1$}\thanks{ Corresponding author: Tel:+353 1 6083165; Fax +353 1 6711759; Email: simon.cox@tcd.ie}}} and {\bf{F. Graner{$^2$}
\thanks{ CNRS UMR 5588 \& Universit{\'{e}} Grenoble I}}}\\
$^1$ {\small Department of Pure and Applied Physics, Trinity
College, Dublin 2, Ireland.}\\
$^2$ {\small Laboratoire de Spectrom{\'{e}}trie Physique,
Bo{\^{\i}}te Postale 87, F-38402 St. Martin d'H{\`{e}}res Cedex,
France.}}

\begin{document}

\maketitle

\begin{abstract}
We consider three-dimensional clusters of equal-volume bubbles 
packed around a central bubble and calculate their energy and optimal shape. We obtain the surface area
and bubble pressures to improve on existing growth laws for three-dimensional
bubble clusters. We discuss the possible number of bubbles that can
be packed around a central one: the ``kissing
problem'', here adapted to deformable objects.
\end{abstract}

Pacs numbers: 82.70.Rr, 83.80.Iz

\section{Introduction}

\subsection{Motivation}
\label{sec:motiv}

Bubbles, such as soap bubbles, are  objects with simple geometry and  physical
properties. But when two or more bubbles cluster together, how
well do we really understand their properties?

The limiting case of a cluster of many bubbles, known as a foam, is 
usually approached with continuum approximations. An understanding of foam properties such as aging, due to gas diffusion, and
structure is a problem of fundamental interest  stimulated  by the need to
predict the behaviour of foam in  industrial applications. From carbonated
drinks to the processes used to extract gold ore from  the earth, foams are  an
important part of our lives with various  industrial uses
\cite[]{WeaireH99,zithabv00}.

The alternative to the continuum description, described here, is an approach
based upon the study  of finite clusters of bubbles. Its advantage is the ease
with which  we can obtain precise structural information. A further benefit of studying finite, rather than infinite or periodic, foams is that the bubbles are not ``frustrated'', so that we get a measure of their free shape, rather than one influenced by long-distance correlations between bubbles.

This has been 
demonstrated convincingly in two-dimensions (2D), where exact results  exist
for two problems of paramount interest:

\begin{itemize}
\item The Kelvin problem: what is the least energy (equivalent to 
line-length) structure of equal-size bubbles that fills space? In 2D, 
\citet{hales01} proved that this is the familiar honeycomb structure. 
In 3D, where the problem is one of minimizing surface energy or area, 
no such exact result exists. Kelvin \cite[]{kelvin87} gave a 
candidate structure, still believed to be the best for a structure 
containing {\em identical} cells, although in the general case it has 
since been beaten by the Weaire-Phelan structure \cite[]{WeaireP94b} 
consisting of bubbles of two different types. The important quantity in
this problem is the surface area of each face of a bubble of unit 
volume, or equivalently the normalized total surface area $S/V^{2/3}$.

\item Growth laws: how does a foam age, or {\em coarsen}, due to gas 
diffusion across its surfaces? The 2D result, due to 
\cite{neumann52}, says that the growth-rate of a bubble (of area $A$) is directly 
linked to its number of sides, $n$: $dA/dt \propto (n-6)$.
That is, it depends upon bubble topology only, irrespective of the precise geometry. In 3D, the
growth law is written  \cite[]{glazier93}:
\begin{equation}
\frac{3}{2 D_{eff}} \frac{\rm{d}V^{2/3}}{\rm{d}t} = 
\frac{1}{2}\Sigma_i \frac{\Delta p_i S_i}{V^{1/3}}
\label{eq:growth1}
\end{equation}
where the sum is taken over each face, which has a pressure difference $\Delta p_i $  and area $S_i$. $D_{eff}$ is an effective diffusion coefficient. 
Again, the normalized area appears to be important, but does the 3D growth law depend only on the bubble topology? In fact it does not, but it may make sense to express the average growth-rate of $F$-faced bubbles as a function of $F$ only {\em if} the dispersion about such a law is small.
\end{itemize}

\subsection{State of the art}
\label{sec:state}

The study of 3D foam coarsening was pioneered by \citet{glazier93}, 
who used a 3D Potts model to numerically simulate foam coarsening. He 
proposed a linear growth law $\Sigma_i \Delta p_i S_i \propto 
(F-cst)$ for bubbles with a number of faces $F$ from 6 to 57 (and 
even from 4 to 60, with some numerical uncertainty).
Similar linear laws were observed in subsequent experiments involving optical tomography and reconstruction 
using the Surface Evolver \cite{monnereauva98} ($F$ between 9 and 16), and  magnetic resonance imaging experiments \cite[]{gonataslygp95,glazierp00} for $F$ from 4 to 26.

This growth law  was refined by  three detailed results presented by 
\citet{hilgenkks01}: first, an approximate analytical formula based upon regular $F$-faced polyhedra with curved faces:

\begin{equation}
\frac{3}{2 D_{eff}} \frac{\dd{\;}}{\dd{t}}V^{2/3}  = G(F) = \frac{3}{2^{1/3}} 
\left[ (F-2) \tan\left(\frac{\pi}{\eta_F}\right) \right]^{2/3} 
\tan^{1/3}\left(\frac{\chi_F}{2}\right) 
\left( \frac{\pi}{3}- \chi_F \right)
\label{eq:hilgengf}
\end{equation}
where $\chi_F = 2 \tan^{-1}\sqrt{4 \sin^2(\pi/\eta_F)-1}$ and $\eta_F = 6-12/F$ is the number of edges per face. For large $F$, this shows a square-root dependence, $G(F \gg 1) = 2.14 \sqrt{F} -7.79$, effective for $F$ greater than about 15. Second, a (non-explicit) correction  for non-regular faces; third,   numerical 
(Surface Evolver) simulations for foams containing bubbles with $F$ 
from 5 to 42.
Recently, \citet{Coxf03} also used  the Surface Evolver to calculate 
numerically the structural properties of single ``regular'' bubbles 
with surfaces of constant mean curvature; this gave
information for certain values of $F$ between 2 and 32.

\subsection{Outline of this paper}
\label{sec:outline}

Here, we study clusters consisting of one bubble surrounded by $F$ 
others, each with prescribed volumes. This constitutes a finite 
cluster with free boundary conditions: this represents a realistic 
foam surrounded by air, in contrast to the idealisation used to derive (\ref{eq:hilgengf}) \citet{hilgenkks01}. We chose such an approach, which neglects
long-distance correlations between bubbles, because it should provide 
 more physical insight than existing experiments and 
simulations, and enable more  precise calculations than the 
analytical approach.
Within this ``mean-field'' choice, all results presented below are 
highly accurate, without approximations. Moreover, in principle we should 
have access to all physically realizable values of $F$.

In the course of our study of the equal-volume case, we encountered
what we call ``the kissing problem for (deformable) bubbles".  Our 
simulations allow us to ask: how many deformable (dry) bubbles can be 
packed around one other?
  The original kissing problem, discussed by Gregory and Newton, was: how many identical hard spheres can surround one 
other, each touching the central one \cite[]{conways99}? In 
two-dimensions the answer is obvious and well-known -- only six hard 
discs can be packed around one other, in the familiar honeycomb 
arrangement. For the three-dimensional problem, consideration of the 
angle subtended by each sphere at the central one suggested that the 
maximum number could be as high as 14, but Newton was correct in 
believing that only 12 neighbours are possible \cite[]{leech56}. We will present arguments suggesting that for bubbles these critical numbers are 12 (2D) and 32 (3D).

The plan of the paper is as follows. We first describe our method of cluster 
preparation and relaxation. There are limits, for each set of given 
bubble volumes, to the values of $F$ for which stable clusters 
exists. In the equal-volume case   we offer a solution to the kissing 
problem. We then analyse in more detail the shape and growth-rate of 
many different bubbles, and present predictions about coarsening and quantify the spread of the growth-rate about the growth law (\ref{eq:hilgengf}).

\section{Definitions and methods}
\label{sec:defns}

We take a central bubble of volume $V_c$ and surround it with $F$ 
bubbles, each with the same volume $V$; this is the natural extension 
into 3D of the 2D ``flower'' of \cite{weairecoxg01}.
To create and equilibrate such a cluster, we use a Voronoi construction with VCS \cite[]{vcs} and then the Surface Evolver 
\cite[]{brakke92}, as follows.

We must first make a choice about the topology of the cluster. Since we wish
 to create the cluster using a Voronoi routine, we must 
first choose an arrangement of $F+1$ points about which to create 
bubbles.

We first place a point at the origin of a sphere of radius 1. Then 
the Voronoi points are placed at the positions given by the solution 
of the ``covering radius problem" \cite[]{conways99}: the arrangement 
of $F$ points on the unit sphere that minimizes the maximum distance 
of any point from its closest neighbour. Candidates to the solution 
of this problem have been given by \cite{hardinss94} for $F$ from 4 
to 130, which is exactly what is required for our purpose. Note that 
this is not the only way to pack the $F$ Voronoi points, but appears 
(partly with hindsight) to have been a good choice -- it gives all 
the arrangements we know to expect, e.g. for $F=6,12,32$.

We truncate the Voronoi diagram by adding $3F$ points at a radial 
distance of $2$ from the origin. We ensure that these outer points 
are at least a distance $2 \epsilon/\sqrt{3F}$ apart, decreasing 
$\epsilon$ from 1  until a solution is found, usually at around 
$\epsilon=0.8$. This data is put through the VCS software; the output 
file is then transferred to the Surface Evolver, version 2.18d. We 
use two levels of refinement and quadratic mode, to obtain a high 
level of accuracy -- we estimate all values to be accurate to at least four decimal places.

We compute the following quantities for the $i^{th}$ face 
($i=1,\cdots, F$) of the central bubble:  its number of sides $n_i$, 
area $S_i$ and pressure difference $\Delta p_i$. Then for the whole 
bubble we record  its volume $V$, its normalized total line length 
$L/V^{1/3}$, its
normalized surface area $S/V^{2/3}$ (where $S = \Sigma_i S_i$), and 
its growth-rate through eq. (\ref{eq:growth1}), which we plot as a 
function of $F$.

\section{Topology and limits for equal-volume clusters}
\label{sec:topology}

\begin{figure}
\centerline{
\includegraphics[width=15cm]{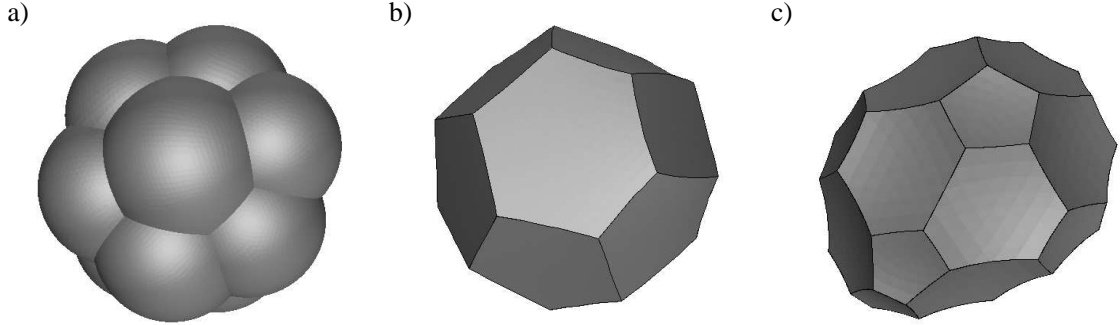}
}
\caption{Examples of the clusters considered here with $V_c = V = 1$ 
in each case: (a) cluster
of $F=13$ outer bubbles (that is, a total of $F+1=14$ bubbles)  still 
attached; (b) the central bubble, drawn to a different scale, with 
$F=13$ faces (this is a Matzke cell \cite{kraynikrvs02} with one square, ten pentagonal and two hexagonal faces); (c) a 
bubble with $F=26$ faces (all pentagonal or hexagonal) -- note its 
departure from approximate sphericity, described in \S 
\ref{sec:kiss}.}
\label{fig:picsVc1}
\end{figure}

We first consider the case where the volume of the central bubble is 
equal to that of its neighbours, $V_c = V$. Examples of such 
monodisperse clusters are shown in figure \ref{fig:picsVc1} for 
$F=13$ and 26. This illustrates that despite the rather symmetric 
initial condition (putting points on a sphere) we can still obtain 
significantly skewed bubbles after relaxation.

\subsection{The kissing problem for 2D bubbles}
\label{sec:kiss2D}

For completeness, we consider first the two-dimensional problem.
How many 2D bubbles can be packed around one other of the same area?

\begin{figure}
\centerline{
\includegraphics[width=15cm]{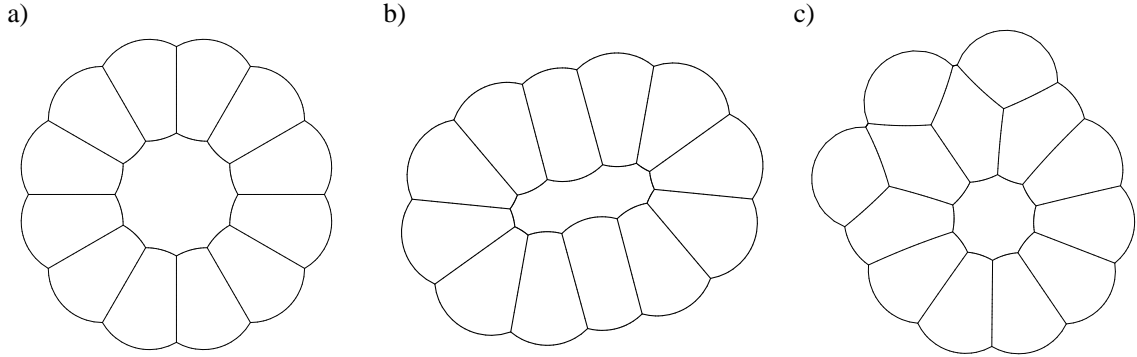}
}
\caption{(a) A symmetrical 2D flower cluster with $F=12$ petals and 
$A_c/A = 2$. (b) One of the possible stable buckled states of the 
same cluster with $A_c = A$. (c) One of the possible ``ejected'' states \cite{coxvw03} with $F=13$ petals and $A_c = A$.}
\label{fig:2dkiss}
\end{figure}

Our initial  pattern is that of the ``flower'' clusters introduced 
recently \cite{weairecoxg01,brakkem02,coxvw03}. It consists of a central cell of area $A_c$ surrounded by $F$ identical petals of area $A$. A 
symmetric  example with $F=12$ petals and $A_c = 2A = 2$ is shown in 
figure \ref{fig:2dkiss}(a).
A priori, one could imagine that the number of petals could  increase 
without limit, with the $F$ sides of the central bubble becoming 
increasingly curved.

However, \citet{weairecoxg01} showed that for $F>6$ there is a ``buckling'' 
instability at a critical ratio of the bubble areas given 
approximately by
\[
A_c/A \approx 0.04(F-6)^2.
\]
For unit areas and $F>6+(0.04)^{-1/2} = 11$, the symmetric shape is 
therefore no longer stable, the flower becomes ``floppy'' and many 
modes of buckling, corresponding to different shaped central bubbles, 
are possible (all with the same energy). An example for $F=12$, in which there is an elliptical mode of buckling, is shown in figure 
\ref{fig:2dkiss}(b).

Is it possible to pack even more bubbles? We find that for $F > 12$, 
any of the buckled configurations of clusters with unit areas are 
unstable to a topological change caused by the length of one of the 
internal edges shrinking to zero \cite{coxvw03}. An example is shown in figure \ref{fig:2dkiss}(c) for $F=13$, for which three bubbles are ``ejected'' in an equilibrium configuration with $A_c/A=1$. 

Thus we conjecture that the maximum number of bubbles that can touch the
central one is 12. This is twice the value for hard discs.

\subsection{The kissing problem for 3D bubbles}
\label{sec:kiss}

In three dimensions the idea is the same. In principle one could 
imagine that there should be no limit to the number of bubbles which 
will fit around the central one, albeit with the latter being hugely 
distorted. However, since the area of each of the five-sided faces 
shrinks as $F$ increases, our simulations of bubbles with unit 
volumes, $V_c/ V = 1$ do not find a stable cluster for all possible
values of $F$.
In fact, we could only find clusters for $5\le F \le 32$.
That is, we cannot obtain a bubble with more than $32$ faces and 
volume equal to that of its neighbours which satisfies Plateau's laws 
after energy (surface area) minimization.

For most values of $F$, the shrinkage of five-sided faces is 
accelerated by an ellipsoidal distortion of the central bubble (see 
figure \ref{fig:picsVc1}(c)), due to the asymmetric location of the 
pentagonal faces amongst the hexagonal ones. Might there be a 
discontinuous buckling transition for 3D clusters? As a result of 
further simulations, we believe not: this asymmetry, and the 
consequent elliptical deformation of the central bubble, means that 
the transition to the asymmetric pattern is continuous.

It is interesting to note that the case   $F=32$ is special: it is 
probably the most symmetric cluster for $F>12$ -- it corresponds to 
the $C_{60}$ fullerene. Hence, by analogy, one might 
expect that stable clusters with unit volumes exist for higher order 
carbon structures. We  tried $C_{80}$ ($F=42$) and the 
elliptical $C_{72}$ ($F=40$) and didn't find them to be stable. We 
thus conjecture that no more than 32 bubbles can touch the central 
one:  32 appears to be the ``kissing'' number for 3D bubbles.
Recall that for hard spheres the kissing number is 12.

\section{Shape, pressure and growth-rate}
\label{sec:spgrv1}

We next analyse in detail the statistics of the bubbles found in our 
simulations.

\subsection{Equal-volume bubble clusters}

\begin{figure}
\psfrag{Surface Area S/V^{2/3}}{\large Surface Area  $S/V^{2/3}$}
\psfrag{Number of faces $F$}{\large Number of faces $F$}
\centerline{
\includegraphics[angle=270,width=15cm]{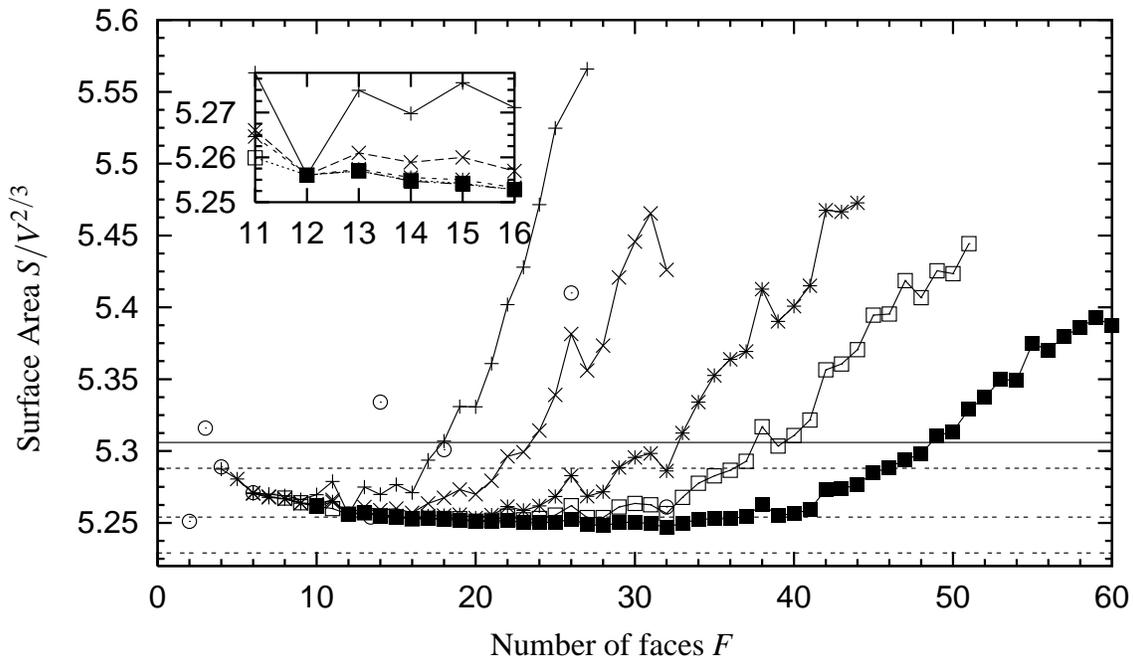}
}
\caption{The normalized surface area $S/V^{2/3}$ varies, albeit over 
a small range, non-monotonically as $F$ increases. Inset: zoom over 
the range $F=11$ to $16$. Data is shown for volume ratios of $V_c/V = \frac{1}{2} (+),1 (\times),2 (+\hspace{-0.33cm}\times),3 (\boxdot)$ and $5 (\blacksquare)$. For all values of the volume ratio $V_c/V$, the 
pentagonal dodecahedron at $F=12$ has the same value of $S/V^{2/3}$, but for all other $F$ the surface area fluctuates widely, although in general it decreases as the volume ratio increases. Also shown is the data for bubbles with constant curvature $(\odot)$ rather than  with fixed
volume \cite{Coxf03}. Shown as horizontal lines (from top to bottom) are the value of $S/V^{2/3}$ for the Kelvin structure (solid line), for the Weaire-Phelan structure (quadruple dashes), for the ``ideal'' flat-faced bubble (triple dashes) and for an infinitely large bubble with hexagonal faces  (double dashes) \cite{kraynikperscomm}.}
\label{fig:coversv23}
\end{figure}

We consider first the  monodisperse case, relevant to the Kelvin 
problem,  where the volume of the central bubble is equal to that of 
its neighbours, $V_c=V$. As mentioned above, we can go from $F$ = 5 
to 32. The ratio $S/V^{2/3}$, shown in figure \ref{fig:coversv23}, is 
lowest at $F=12$ (consistent with data for random monodisperse foams \cite{kraynikrvs02}), and increases steeply for $F$ greater than about 16.

The inset on figure \ref{fig:coversv23} shows the data around the 
optimal region $F=11$ to 16. These bubbles, which do not pack to 
fill space, have lower area than Kelvin's (5.306) and even Weaire-Phelan's 
(5.288) (see \cite{kraynikrvs02} for details of other space-filling foam structures). They are barely above the value for the so-called ``ideal'' 
bubble (5.254) \cite[]{isenberg92}. The latter, with $F= 13.39$, describes a regular (but unphysical) ``bubble'' which would have flat faces, and hence a growth-rate of zero.

\begin{figure}
\psfrag{Edge length L/V^{1/3}}{\large Line length $L/V^{1/3}/F^{1/2}$}
\psfrag{L/V^{1/3}}{\large $L/V^{1/3}$}
\psfrag{Number of faces $F$}{\large Number of faces $F$}
\centerline{
\includegraphics[angle=270,width=15cm]{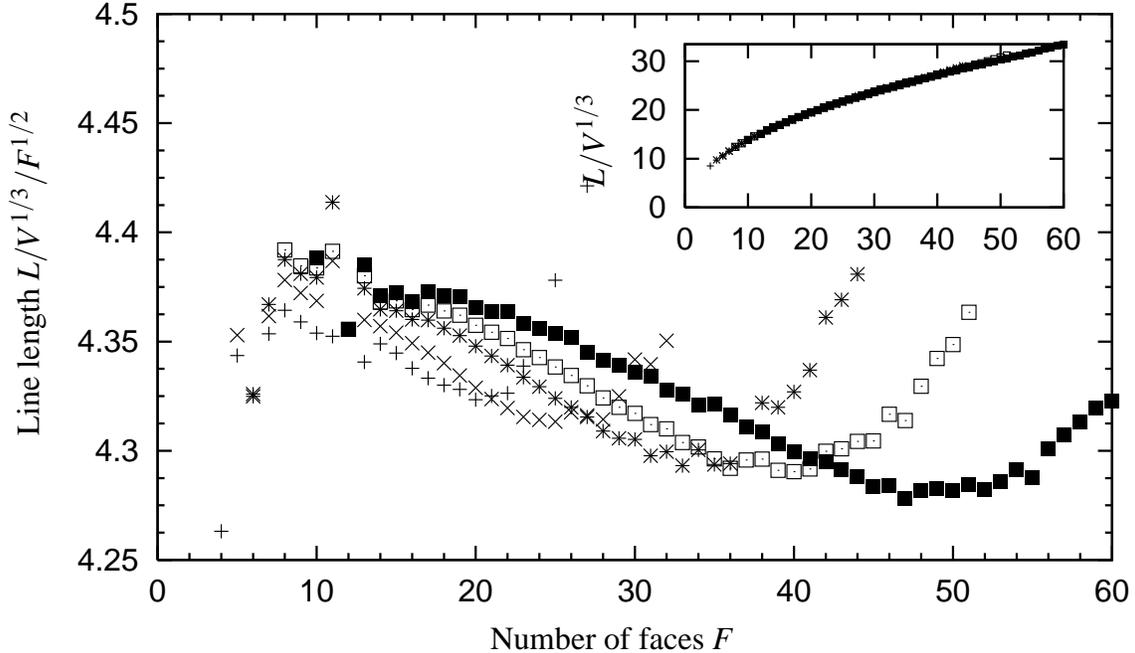}
}
\caption{The line-length ratio $L/V^{1/3}$ increases in proportion to the square-root of $F$ \cite{Coxf03} (inset). We therefore plot the ratio $L/V^{1/3}/F^{1/2}$, for volume ratios of $V_c/V = \frac{1}{2} (+),1 (\times),2 (+\hspace{-0.33cm}\times),3 (\boxdot)$ and $5 (\blacksquare)$. The data is everywhere close to $L = 4.35 V^{1/3} F^{1/2}$, confirming the square-root behaviour. The dispersity increases as both $F$ and $V_c/V$ increase.}
\label{fig:coverlv13}
\end{figure}

Also of interest is the normalized line-length $L/V^{1/3}$ of each 
bubble, plotted in figure \ref{fig:coverlv13}. Note that all data 
lies close to a line $L/V^{1/3} \propto \sqrt{F}$ \cite{Coxf03}. We therefore show the ratio $L/V^{1/3}/F^{1/2}$ in figure \ref{fig:coverlv13}: the maximum deviation  (i.e. the shallow minimum in the data) occurs for $F \approx 25$.

\subsection{Non-equal volumes}

\subsubsection{Simple volume ratios}

We next consider the case where the volume of the central bubble is  not equal
to the volume of its neighbours. There are again limits to the possible stable
clusters, but they vary with  the volume ratio. We study the simple ratios
$V_c/V =  \frac{1}{2},2, 3 $ and $5$. This choice of volume ratios allows us to
explore $F$ from 4 to 60.

Note that  the possible range of $F$ is not always continuous. For instance, we
cannot  construct a stable cluster with 26 neighbours for $V_c/ V = 
\frac{1}{2}$, hence  we find $F \in [4-25,27]$. Similarly, 11 neighbours is
unstable for $V_c/ V = 5$, and we find $F \in [10,12-60]$.

For each value of $F$ we record the topology of each bubble, collated 
for all volume ratios (table \ref{tab:face_sides}), using the 
notation $n_x$ to mean that the bubble has $x$-faces with $n$-sides. 
The topology of the central bubble might depend on $V_c/V$:  we find 
such  non-uniqueness in only two instances.
We accept this as due to the slight randomness used in placing the 
$3F$ outer points to truncate  the initial Voronoi pattern.

\begin{table}
\caption{The topology of each central bubble, where $n_x$ denotes the 
number $x$ of $n$-sided faces. 
\newline $\;^\ast$ denotes configurations for $F=2$ and 3 from \cite{Coxf03}. 
\newline $\;^{\ast\ast}$ denotes alternative
configurations for given $F$ with different $V_c/V$: $F= 11$ ($4_2 
5_8 6_1$ for $V_c/V = 3$) and $34$ ($5_{14} 6_{20}$ for $V_c/V = 2$). 
}
\label{tab:face_sides}
\begin{center}
\[
\begin{array}{|c||c||c||c|}\hline
   \begin{array}{c|c}
   F   & \mbox{Topology} \\ \hline\hline
   1 &-\\
   2 &{1_2}^{\ast} \\
   3 &{2_3}^{\ast}  \\
   4   & 3_4 \\
   5   & 3_2 4_3 \\
   6   & 4_6 \\
   7   & 4_5 5_2 \\
   8   & 4_4 5_4 \\
   9   & 4_3 5_6\\
  10   & 4_2 5_8  \\
  11   & {4_3 5_6 6_2}^{\ast\ast} \\
  12   & 5_{12}  \\
  13   & 4_1 5_{10} 6_2 \\
  14   & 5_{12} 6_2 \\
  15   & 5_{12} 6_3
  \end{array} &
  \begin{array}{c|c}
   F   & \mbox{Topology} \\ \hline\hline
  16   & 5_{12} 6_4 \\
  17   & 5_{12} 6_5 \\
  18   & 5_{12} 6_6 \\
  19   & 5_{12} 6_7 \\
  20   & 5_{12} 6_8 \\
  21   & 5_{12} 6_9 \\
  22   & 5_{12} 6_{10} \\
  23   & 5_{12} 6_{11} \\
  24   & 5_{12} 6_{12} \\
  25   & 5_{12} 6_{13} \\
  26   & 5_{12} 6_{14} \\
  27   & 5_{12} 6_{15} \\
  28   & 5_{12} 6_{16} \\
  29   & 5_{12} 6_{17} \\
  30   & 5_{12} 6_{18}
  \end{array} &
  \begin{array}{c|c}
   F   & \mbox{Topology} \\ \hline\hline
  31   & 5_{13} 6_{17} 7_1 \\
  32   & 5_{12} 6_{20} \\
  33   & 5_{13} 6_{19} 7_1 \\
  34   & {5_{12} 6_{22}}^{\ast\ast} \\
  35   & 5_{14} 6_{19} 7_2 \\
  36   & 5_{14} 6_{20} 7_2 \\
  37   & 5_{12} 6_{25}\\
  38   & 5_{12} 6_{26} \\
  39   & 5_{12} 6_{27} \\
  40   & 5_{12} 6_{28} \\
  41   & 5_{12} 6_{29} \\
  42   & 5_{12} 6_{30} \\
  43   & 5_{12} 6_{31} \\
  44   & 5_{12} 6_{32} \\
  45   & 5_{13} 6_{31} 7_1
  \end{array} &
  \begin{array}{c|c}
   F   & \mbox{Topology} \\ \hline\hline
  46   & 5_{12} 6_{34} \\
  47   & 5_{14} 6_{31} 7_2 \\
  48   & 5_{12} 6_{36} \\
  49   & 5_{12} 6_{37} \\
  50   & 5_{12} 6_{38} \\
  51   & 5_{12} 6_{39} \\
  52   & 5_{13} 6_{38} 7_1 \\
  53   & 5_{13} 6_{39} 7_1  \\
  54   & 5_{12} 6_{42} \\
  55   & 5_{14} 6_{39} 7_2  \\
  56   & 5_{12} 6_{44} \\
  57   & 5_{12} 6_{45} \\
  58   & 5_{12} 6_{46} \\
  59   & 5_{12} 6_{47} \\
  60   & 5_{12} 6_{48}
   \end{array}
  \\ \hline
  \end{array}
\]
\end{center}

\end{table}

The line-length, shown in figure \ref{fig:coverlv13}, fall close to the same 
curve as in the monodisperse case.
The square-root approximation becomes slightly worse as the bubbles become larger and gain more faces, with the maximum deviation occurring at higher $F$ for increasing $V_c/V$.

\subsubsection{Large  volume ratios}

With larger $V_c/V$ we can look at bubbles with many faces and very 
low surface areas. For instance, with $F=122$ (corresponding to the 
fullerene $C_{240}$) and $V_c/V = 200$ we find a bubble with topology 
$5_{12}6_{110}$ and $S/V^{2/3} = 5.239$: see figure \ref{fig:optimalsf}.

We  could extend this process to larger bubbles with more faces. The 
normalized area should eventually approach the value for an 
infinitely large bubble with hexagonal faces: $S/V^{2/3} = 5.229$ 
\cite[]{kraynikperscomm}; note that this is not the
theoretical lower bound for the normalized area, which corresponds to 
a  spherical bubble with $F-1$ infinitesimally small neighbours 
\cite[]{brakkeperscomm}.

\subsubsection{Correlations}

Real foams often have a distribution of bubble volumes, and their 
topology  is correlated  to the geometry: larger bubbles tend to have 
more neighbours \cite[]{weaireg93}.

\begin{figure}
\psfrag{Surface Area S/V^{2/3}}{\large Surface Area  $S_{\mathrm{opt}}/V^{2/3}$}
\psfrag{Number of faces $F$}{\large Number of faces $F$}
\psfrag{Optimal data}{\large Small $V_c/V$}
\psfrag{Large cluster}{\large Large $V_c/V$}
\psfrag{Lower bound}{\large Limiting value}
\psfrag{Power-law fit}{\large Power-law fit}
\centerline{
\includegraphics[angle=270,width=15cm]{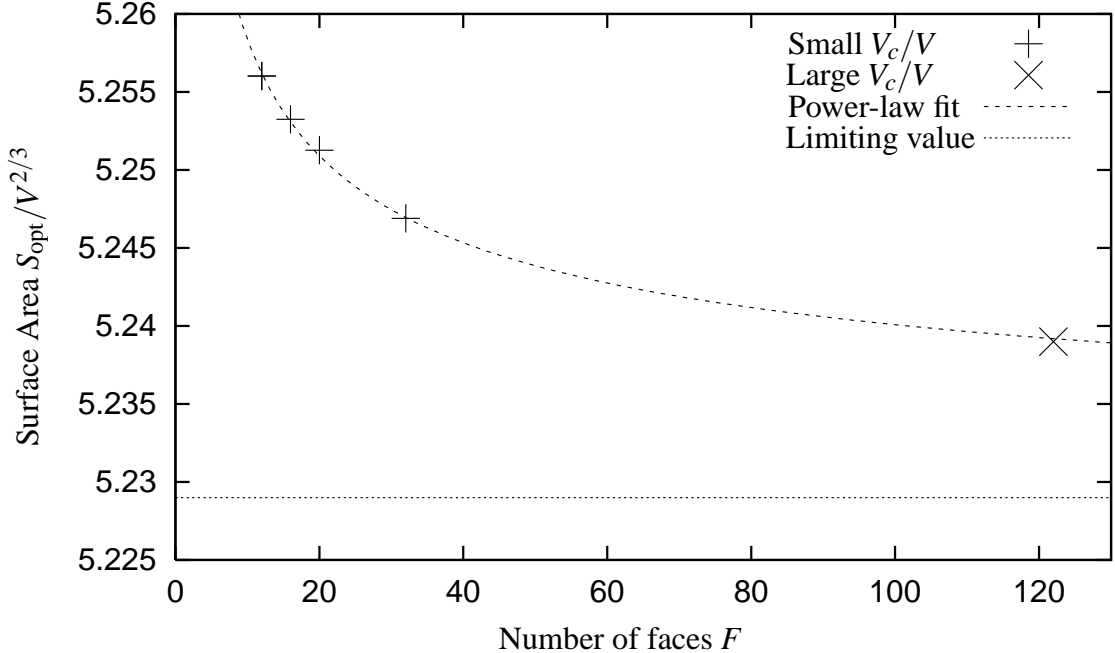}
}
\caption{The optimal normalized surface area $S_{\mathrm{opt}}/V^{2/3}$ for a range of values of the volume ratio $V_c/V$. Data are shown for values of $V_c/V = 1,2,3$ and $5$ ($+$) and for the representative calculation for $F=122$ with $V_c/V=200$ ($\times$). The limiting value for $V_c/V \rightarrow \infty$ at $S/V^{2/3}= 5.229$ is shown as a horizontal line and we also show a power-law fit: $S/V^{2/3}= 5.229 + 0.078/F^{0.423} $.}
\label{fig:optimalsf}
\end{figure}

Such correlations appear in our results, although we did not 
specifically include them. Their physical origin is clear. In fact, 
consider a bubble of volume $V_c$, and
consider the  average of its neighbours' volumes, denoted $V$ 
(mean-field description).
Then, for this given $V_c/V$ ratio, the physically realizable values 
of $F$ are limited. Within the possible $F$, the $S(F)$ curves admit 
an optimum: there is a value of $F$ which minimizes the bubble area.
These optimal $F$ values do increase with $V_c/V$.
Moreover, on figure \ref{fig:coversv23} we can read the optimal 
surface  $S_{\mathrm{opt}}/V^{2/3}$ as a function of $F$: it is the 
envelope of all curves plotted, shown in figure \ref{fig:optimalsf}. It decreases roughly as one over the square-root of $F$ as the volume ratio increases.

In 2D, the expression for $L_{\mathrm{opt}} /A^{1/2}$ {\it versus} $n$ has been
used to estimate the energy of a 2D foam \cite[]{granerjjf01}, then to
determine the correlations between geometry (area $A$) and topology (number of
sides $n$) \cite[]{fortest03}. Here, its  3D counterpart, the $S_{\mathrm{opt}}
/V^{2/3}$ {\it  versus} $F$  relation, is more complicated (in particular,
unlike in 2D, it depends on the volume ratio) \cite{kraynikperscomm}: but it
appears to have the same essential property as in 2D, namely to be a
non-increasing function of  $F$; we thus hope to extend to 3D this 2D result
\cite{fortest03}.

\begin{figure}
\psfrag{Scaled Volume V/L^3}{\large Scaled Volume $V/\hat{l}^3$}
\psfrag{Scaled area S/L^2}{\large Scaled area $S/\hat{l}^2$}
\psfrag{Number of faces $F$}{\large Number of faces $F$}
\centerline{
\includegraphics[angle=270,width=15cm]{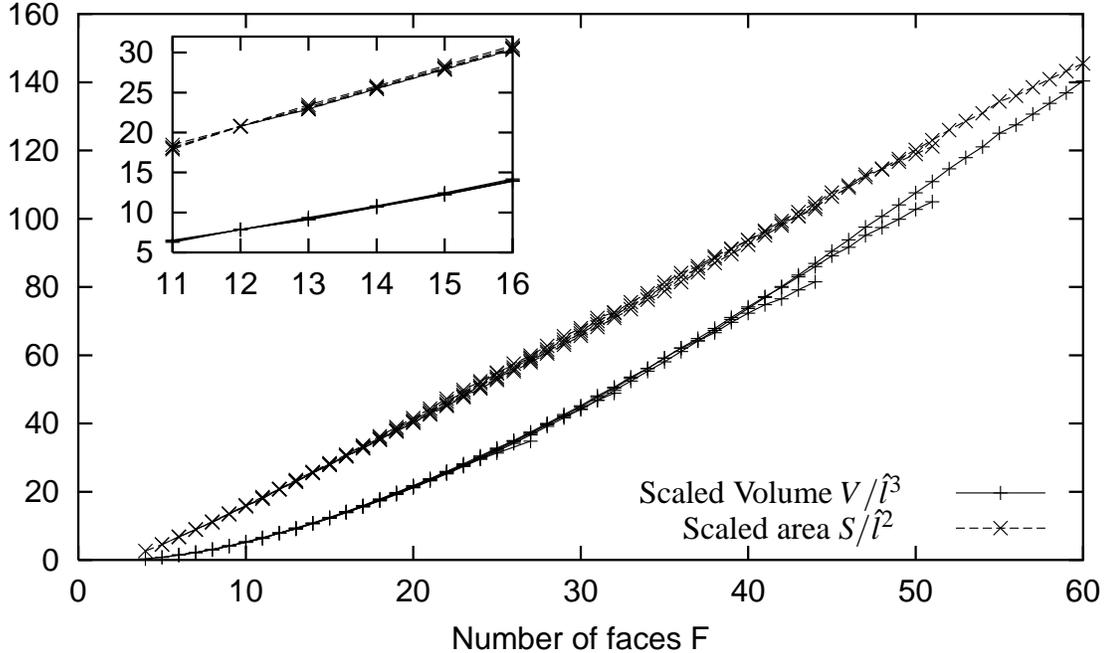}
}
\caption{The volume $V$ and surface area $S$ of a bubble with $F$ faces scaled with the average length of an edge $\hat{l}$. The data for the surface area is  affine, $S/\hat{l}^2 = 2.573 F -9.801$, while for low $F$ the volume data is approximately quadratic, $V/\hat{l}^3 \approx 0.053 F^2$. Data is shown for all volume ratios $V_c/V$ considered here, and the inset shows details of $V$ and $S$ for bubbles with 11 to 16 faces.}
\label{fig:vsl}
\end{figure}

In the theory of foam drainage, in which liquid flows along the
edges separating the faces (Plateau borders), and the coupling of drainage with
coarsening, it is useful to know the following two dimensionless parameters
\cite{WeaireH99,hilgenks00b}: $V/\hat{l}^3$ and $S/\hat{l}^2$, where $\hat{l}$
is the average length of an edge in an $F-$faced bubble. We can calculate these
quantities from our results, and they are shown in figure \ref{fig:vsl}; both
increase strongly with the number of faces $F$ and are insensitive to the size of the neighbouring bubbles.

\subsection{Growth-rate}

\begin{figure}
\psfrag{dV23/dt}{\large ${\rm{d}V^{2/3}}/{\rm{d}t}$}
\psfrag{dV23/dt-G(F)}{\large ${\rm{d}V^{2/3}}/{\rm{d}t}-G(F)$}
\psfrag{Number of faces $F$}{\large Number of faces $F$}
\centerline{
\includegraphics[angle=270,width=15cm]{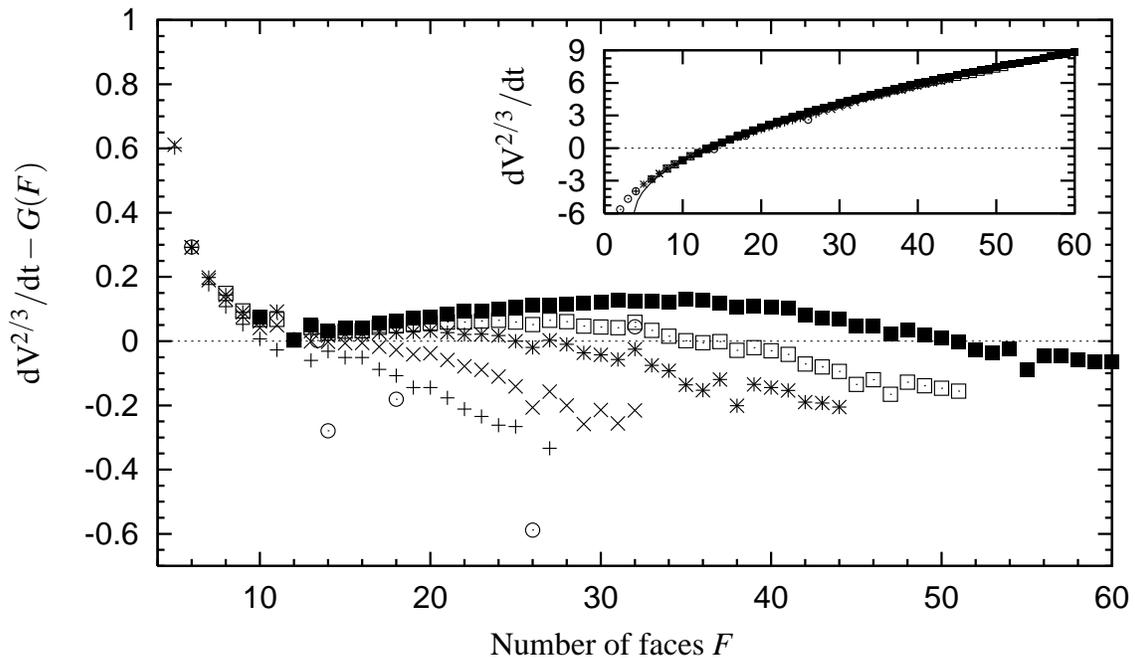}
}
\caption{The difference in the rate of change of volume of a bubble with $F$
faces, calculated from our simulations using (\ref{eq:growth1}), and the value calculated according to
the analytic formula (\ref{eq:hilgengf}). The inset shows the values
themselves, again next to the analytic line, from which it deviates at small
$F$. Data is shown for volume ratios of $V_c/V = \frac{1}{2} (+),1 (\times),2
(+\hspace{-0.33cm}\times),3 (\boxdot)$ and $5 (\blacksquare)$. The data for bubbles with constant curvature $(\odot)$ \cite{Coxf03}, rather than fixed
volume, is more scattered, but useful for low $F$.}
\label{fig:coverdvdt}
\end{figure}

As a result of these simulations, we are able to calculate the instantaneous growth-rate of many bubbles, with many different numbers of sides, through the formula (\ref{eq:growth1}). It is shown in the inset to figure \ref{fig:coverdvdt} -- all data lies close to (\ref{eq:hilgengf}), except at (for us unobtainable) small $F$ where the results of \citet{Coxf03} are useful.

More instructive is the difference between the analytic formula and 
our data, shown in figure
\ref{fig:coverdvdt}.
For $F \ge 12$, our data are above and below the analytic line: it agrees 
with the suggestion that the analytic formula approximates the 
average growth-rate \cite{hilgenkks01}, and quantifies the dispersion 
around this average (less than 1 \% dispersion). Conversely, for 
$F<12$, our data are clustered and significantly (up to 10 \%) larger 
than the analytical formula, which confirms that the analytical 
approximations gradually lose their validity at low $F$, as expected 
\cite{hilgenkks01}. In a coarsening foam, the bubbles with low $F$ are important because it is these bubbles that disappear. So although for $F\ge 12$ the growth-rate is well approximated by (\ref{eq:hilgengf}), we give in table \ref{tab:lowfgrow} the growth-rates for bubbles with $F<12$, averaged over all simulations. These are almost 
indistinguishable from the growth rates calculated exactly on ideal 
bubbles \cite{kraynikperscomm}: the agreement 
for $F=4$ to $11$ is better  than 1\%, and it is even lower than 0.1\% 
for regular bubbles ($F=4$ and 6).

\begin{table}
\caption{The growth-rates, averaged over all simulations, for bubbles with few faces, $F<12$. They differ significantly from the analytic equation (\ref{eq:hilgengf}) \cite{hilgenkks01}, but show very little dispersion.}
\label{tab:lowfgrow}
\begin{center}
\[
  \begin{array}{|l|c|c|c|c|c|c|c|c|c|c|}\hline
   F                 & 2 & 3 & 4 & 5 & 6 & 7 & 8 & 9 & 10 & 11 \\ \hline
  -{\rm{d}V^{2/3}}/{\rm{d}t} & 5.632 & 4.655 & 3.967 & 3.326 & 2.849 & 2.350 & 1.899 & 1.506 & 1.130 & 0.760 \\ \hline   
   \end{array}
\]
\end{center}

\end{table}

\section{Conclusions}

The structure of a foam in equilibrium minimizes its (free) energy, which is
the product of (i) two quantities characterising the system (surface tension
and average surface area) and (ii) some function of shape only. The structure
changes due to coarsening. The coarsening rate is the product of a diffusion
constant (which depends on the material parameters, including chemical
composition), that sets the characteristic time scale, and a function only of
geometry. Here, we don't consider two other phenomena, drainage and film
breakage, which cause deviations from equilibrium. 

Using the Surface Evolver, we have studied finite clusters of bubbles to give
information about the structure of three-dimensional foam and a 3D coarsening
law. This approach allows us to get a high level of detail and accuracy of
the  relevant quantities (surface area, pressure difference) to get a good 
insight into how foams coarsen. Our calculated values of the growth law  require
no assumption about the curvature being small, and can be found for bubbles
with an arbitrary number of faces.

As the volume ratio between the central bubble and its neighbours 
changes, we find upper and lower bounds on the possible number of 
faces, because the bubbles deform. This leads us to conjecture a 
value for the kissing problem for foams: no more than 32 bubbles can 
be stably packed around one other of the same volume.

Although we don't tackle infinite (or, equivalently, periodic)
structures, we expect that this data will eventually lead to greater
insight into the Kelvin problem, since we are starting to understand 
better what happens for bubbles with between 12 and 16 faces.

\section*{Acknowledgements}

We thank Professor K. Brakke for having distributed and maintained  his Surface
Evolver program. This work benefited from a stay at the  Isaac Newton Institute
in Cambridge, and, in particular, discussions with A. Kraynik, S. Hilgenfeldt and J. Glazier. Financial support is gratefully acknowledged from the  Ulysses France-Ireland exchange scheme.

\end{document}